# A New Cumulant Expansion Based Extraction for Higher Order Quantum Corrections in Equilibrium Wigner-Boltzmann Equation

Xiyue Li[1, a)] and Everett X. Wang[1, b)]

[1]*School of Information Engineering, Guangdong University of Technology, Guangzhou, China*

Email：a）lixiyue915@gmail.com

b）everett.wang@gdut.edu.cn

**ABSTRACT**：A Cumulant based method has been introduced to extract quantum corrections in distribution function with the equilibrium Wigner-Boltzmann equation. It is shown that unlike the moment expansion used in hydrodynamic model, cumulant expansion converges much faster when distribution function is closed to Maxwellian with only first three cumulants are non-zero. In this case, quantum corrections higher than first order can be extracted mainly by lowest three cumulants and odd number derivatives of potential function. This method also provides a new way to determine the distribution function with Maxwellian form and study the role of potential function in quantum correction field.

## I. Introduction

In recent years, quantum effects are becoming increasing noticeable as semiconductor devices scaled into the nanometer regime[1-8]. The behavior of quantum effects near thermal equilibrium can be approximated by adding quantum corrections $O(\hbar^2)$, $O(\hbar^4)$ … to the classical transport models. Various methods were applied to include quantum effects in transport simulations, having their respective strengths as well as their limitations. The full three-dimensional (3-D) quantum hydrodynamic model based on moment equation was first derived by Garner[9], in which quantum corrections to macroscopic quantities such as stress tensor and energy density were included. However, only $O(\hbar^2)$ was presented. A quantum model which described electron transport property in graphene, based on Liouville-von Neumann equation and minimum entropy principle, was proposed by Zamponi[10], but they employed many of the same approximation which usually considered in classical analogue. Moreover, distribution function including quantum components did not provided in their work. Bose and Janaki obtained an exact solution for one-dimensional (1-D) Wigner-Boltzmann equation only up to $O(\hbar^2)$[11]. They also expanded the solution be appropriate for all higher order quantum corrections[12], but their format was too complicated to be used for real simulations. In general, the above-mentioned methods cannot provide the concise formula form of distribution function which taken quantum corrections higher than first order into account.

In order to extract the accurate description of distribution function and to reduce the complexity of the calculation, it is believed that cumulant expansion can be regarded as a more appropriate method[13]. The standard hydrodynamic models are based on moment expansion of the semi-classic Boltzmann or quantum Wigner transport equations. The expansion is truncated at the third moment that minimizes entropy function. The truncation makes the problem solvable. But it also introduces errors. In this paper we intend to use cumulant expansion to address above problem since it converges much faster so truncation error is greatly reduced.

Cumulant expansion[14] has been used in many fields of statistics. It is especially efficient in treating random process whose distribution function is close to Maxwellian. We will show that only three cumulants are needed to represented a Shifted Maxwellian instead of the infinite number of terms required by moment expansion. In addition, the higher order cumulant should be small and converge fast even when the distribution functions are distorted from a Shifted Maxwellian.

In this work, cumulant expansion are applied to 1-D Wigner-Boltzmann equation to obtain the quantum corrections of higher than first order with a Shifted Maxwellian distribution function in Section 2. Next, relationship between moment and cumulant are shown in Section 3, by comparing the coefficients of the variables of random momentum. Furthermore, cumulant expansion-based solution for single particle distribution is obtained. And finally, we present the conclusion and future work.

## II. Extraction of the Quantum Corrections

The obtaining of partition function in classical Boltzmann equation is concerned with the determination of energy levels by solving the collision-less Schrodinger equation. The equilibrium Wigner-Boltzmann equation is written as[9]

$$\frac{\partial f}{\partial t} + \frac{p}{m}\frac{\partial f}{\partial x} - \frac{2}{\hbar}\sum_{r=0}^{\infty}(-1)^r \frac{(\hbar/2)^{2r+1}}{(2r+1)!}\frac{\partial^{2r+1}V}{\partial x^{2r+1}}\frac{\partial^{2r+1}f}{\partial p^{2r+1}} = 0 \tag{1}$$

where $f = f(x,p,t)$ is quantum single particle distribution function depending on space coordinate $x$, momentum $p$ and time $t$. $V$ is the average potential. The quantum corrections can be described as a function associated with the Plank constant $\hbar$. The moment expansion of Wigner-Boltzmann equation involves integrating powers of momentum p over distribution function $f(x,p,t)$ in Eq. (1) to obtain the conservation laws for particle number, momentum, energy and higher order moments. Integrating a function $<p^r>$ (r=0,1,2…) with respect to $p$ against the Wigner-Boltzmann equation, we obtain

$$\frac{\partial <p^r>}{\partial t} + \frac{1}{m}\frac{\partial <p \cdot p^r>}{\partial x} - \frac{2}{\hbar}\sum_{r=0}^{\infty}(-1)^r \frac{(\hbar/2)^{2r+1}}{(2r+1)!}\frac{\partial^{2r+1}V}{\partial x^{2r+1}}\frac{\partial^{2r+1}<p^r>}{\partial p^{2r+1}} = 0 \tag{2}$$

Here we set $<p^r>$ with r= 1, 3, 5, yield the moment equations with no quantum correction and $O(\hbar^2)$ and $O(\hbar^4)$ which can be seen in Eqs. (3)-(5), respectively.

$$\frac{\partial <p>}{\partial t} + \frac{1}{m}\frac{\partial}{\partial x}<p^2> + n\frac{\partial V}{\partial x} = 0 \tag{3}$$

$$\frac{\partial <p^3>}{\partial t} + \frac{1}{m}\frac{\partial}{\partial x}<p^4> + 3<p^2>\frac{\partial V}{\partial x} - (\frac{\hbar}{2})^2 n\frac{\partial^3 V}{\partial x^3} = 0 \tag{4}$$

$$\frac{\partial <p^5>}{\partial t} + \frac{1}{m}\frac{\partial}{\partial x}<p^6> + 5<p^4>\frac{\partial V}{\partial x} - (\frac{\hbar}{2})^2 \frac{5!}{2!3!}<p^2>\frac{\partial^3 V}{\partial x^3} + (\frac{\hbar}{2})^4 n\frac{\partial^5 V}{\partial x^5} = 0 \tag{5}$$

From Eqs. (2)-(5), we can derive the form of the quantum corrections and the first order is $O(\hbar^2)$.

$$\sum_{m=3,5,7...(odd)}(\frac{i\hbar}{2})^{m-1}\frac{N!}{(N-m)!m!}<p^{N-m}>\frac{\partial^m V}{\partial x^m} \tag{6}$$

where $N$ and $m$ are integers with $N>m$. Eq. (6) represents the quantum corrections for moments of order $N$. In the limit $\hbar \to 0$, the quantum correction will be vanished and Eq. (2) reduces to the classical equilibrium Boltzmann equation.

When the system is in the stationary state, we have $\partial f/\partial t = 0$. If working in one dimension, the following trial solution in Eq. (1) can be inserted as[11,15]

$$f(x,p,t) = C\exp[-\frac{\beta}{\theta}\frac{(p-p_0)^2}{2}] \tag{7}$$

where $\theta$ are considered to the function of $x$, $\beta$ is inhomogeneous temperature, $p_0$ is the average momentum and $C$ is the appropriate normalization parameter. Using the moment expansion with the distribution function with a

Shifted Maxwellian form, the first, second, third order quantum corrections can be extracted without the partial derivative against time in Eqs. (3)-(5). Higher order quantum corrections can be also extracted with the corresponding order moment. We can see that quantum corrections can be described by moments, and such moments can also be easily calculated by the lowest three cumulants, which is shown in the Next Section.

## III. Cumulant Expansion of Maxwellian Distribution

One popular approach to solve the Wigner-Boltzmann equation is by expanding the distribution function into a series of known base functions. The faster the series converges, the less error is produced due to the closure. Both the cumulant and moment expansions connect closely with the characteristics function. Characteristics function $M(\xi)$ is define as the Fourier transform of its distribution function, which can be obtained as

$$M(\xi) = \sum_{r=0} \frac{f_r \xi^r}{r!} = \int_{-\infty}^{+\infty} [\exp(\xi p) \cdot f] dp \tag{8}$$

where $f_r$ are moment $<p^r>$, which defined as the coefficients of a Taylor expansion of the characteristics function, which can be obtained as

$$f_r = <p^r> = \int (p^r \cdot f) dp \tag{9}$$

On the other hand, cumulants $c_r$ are the coefficients of a Taylor expansion of the logarithm of the characteristics function, with the form of

$$K(\xi) = \log M(\xi) = \sum_{r=0} \frac{c_r \xi^r}{r!} \tag{10}$$

The benefit of cumulant expansion is obvious because only the first three terms in the expansion are needed to represent a Shifted Maxwellian distribution function. Further, as long as the distribution are close to Shifted Maxwellian, the higher order cumulants will be small and reduce to zero quickly. In contrast, the moments of the Shifted Maxwellian distribution function, which are described by Eq. (9), never go to zero until the number of moments trends to infinity.

Eq. (7) is used as to show the effectiveness of the cumulant expansion, it is easy to find that the first three cumulants are $\ln(n)$, $p_0$ and $\theta/\beta$, respectively, all other higher order cumulants are 0. If the distribution function is expanded in moments, the series will need an infinite number. Figure 1 shows the comparison between cumulants and moments series of a Shifted Maxwellian function. The moments increase geometrically with the increasing order, while cumulants are all zero for orders higher than 2.

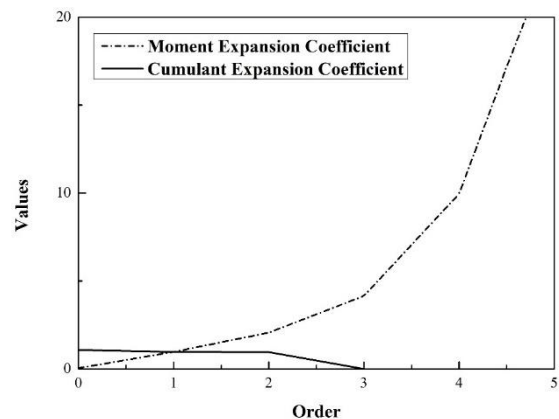

FIG. 1. Comparison between cumulant and moment from a Shifted Maxwellian function

With the interaction $\exp\sum_{r=0}\frac{c_r\xi^r}{r!} = \sum_{r=0}\frac{f_r\xi^r}{r!}$, and evidently $f_0=n$ refer to the carrier concentration, which also implies $c_0=\ln(n)$, the relationships between first few moments and cumulants, obtained by extracting coefficients from the expansion, can be seen as follow,

First order: $n \cdot c_1 = f_1$

Second order: $n(c_1^2 + c_2) = f_2$

Third order: $n(c_1^3 + 3c_1c_2) = f_3$

Fourth order: $n(c_1^4 + 6c_1^2c_2 + 3c_2^2) = f_4$

Fifth order: $n(c_1^5 + 15c_1c_2^2 + 10c_1^3c_2) = f_5$

Sixth order: $n(c_1^6 + 15c_2^3 + 45c_1^2c_2^2 + 15c_1^4c_2) = f_6$

…….

We can see that for every moment, not matter how high the order is, can be calculated by the lowest three non-zero cumulants, in that case we can quickly determine the value for quantum corrections for every order, which can be seen in Eq. (6).

## IV. Extraction for Higher Order Quantum Corrections

In this section, the way to extract the second order quantum corrections which associated with $O(\hbar^4)$ is shown as an example. First of all, we calculate the classical component. When $f_2$ is substituted into Eq. (3) for order $r=2$, we obtain

$$f_2 = n\frac{\theta}{\beta} + np_0^2 \tag{11}$$

$$\frac{n}{\beta m}\frac{d}{dx}\theta_0 = -n\frac{dV}{dx} \tag{12}$$

Solving Eq. (12), the classical component $\theta_0$ can be obtained as a potential dependence function.

$$\theta_0 = -\beta mV \tag{13}$$

Next, in order to obtain the $O(\hbar^4)$ quantum correction, $f_6$, and $f_4$ can be calculated by Eq. (9) and the results are

$$f_4 = 3n\frac{\theta^2}{\beta^2} + 6np_0^2\frac{\theta}{\beta} + np_0^4 \tag{14}$$

$$f_6 = 15n\frac{\theta^3}{\beta^3} + 45np_0^2\frac{\theta^2}{\beta^2} + 15np_0^4\frac{\theta}{\beta} + np_0^6 \tag{15}$$

The same results can be also obtained by the cumulants calculating. Then, substituting $f_6$, $f_4$ and $f_2$ in Eq. (5) without the partial derivative against time, we obtain

$$\begin{aligned}&(\frac{45n}{\beta^3 m}\theta_2^2 + \frac{90np_0^2}{\beta^2 m}\theta_2 + \frac{15np_0^4}{\beta m})\frac{\partial\theta_2}{\partial x} + 5(3n\frac{\theta_2^2}{\beta^2} + 6np_0^2\frac{\theta_2}{\beta} + np_0^4)\frac{\partial V}{\partial x} \\ &-10(\frac{\hbar}{2})^2(n\frac{\theta_2}{\beta} + np_0^2)\frac{\partial^3 V}{\partial x^3} + (\frac{\hbar}{2})^4 n\frac{\partial^5 V}{\partial x^5} = 0\end{aligned} \tag{16}$$

where $\theta_2$ is $\theta_0 + \Delta\theta_2$ and $\Delta\theta_2$ is the second order quantum correction associated with the $O(\hbar^4)$. Substituting $\theta_0$, $\frac{\partial \theta_0}{\partial x}$ in Eq.(16) and assuming $\frac{\partial \Delta\theta_2}{\partial x}$ to be the perturbation, we can finally obtain $\theta_2$ with the form of below.

$$\theta_2 = -\beta mV + \frac{(p_0^2 mV - \frac{1}{2}m^2V^2 - \frac{1}{6}p_0^4)\frac{\partial V}{\partial x} - (\frac{\hbar}{2})^2 \frac{1}{6}(p_0^2 - mV)\frac{\partial^3 V}{\partial x^3} + (\frac{\hbar}{2})^4 \frac{1}{60}\frac{\partial^5 V}{\partial x^5}}{\frac{1}{\beta}[(p_0^2 - mV)\frac{\partial V}{\partial x} + (\frac{\hbar}{2})^2 \frac{1}{6}\frac{\partial^3 V}{\partial x^3}]} \quad (17)$$

$\Delta\theta_2$ is the second part of the right side, which has the correlation with first, third and fifth order derivative of potential. Since every higher order moment can be easily described by the lowest three cumulants when the distribution function is close to Shifted Maxwellian, this process can be continued for $O(\hbar^6)$, $O(\hbar^8)$ ..., from third, fourth order to every higher order quantum correction by determining the $\Delta\theta_3$, $\Delta\theta_4$ ..., respectively.

**V. Conclusion**

In this work, a cumulant expansion-based solution for equilibrium Wigner-Boltzmann equation, valid with quantum corrections higher than the first order, is proposed. First, the description of quantum corrections is extracted, which is associated with odd number derivates of the potential and moments. Second, comparison between cumulant and moment expansions is presented, proving that the lowest three cumulant can represent every higher order moment when distribution function is close to Shifted Maxwellian, which offers convergence advantages over moment expansion-based method such as traditional hydrodynamic model. Third, extraction for quantum corrections higher than the first order in equilibrium Wigner-Boltzmann equation is derived, it is concluded that the quantum corrections can be treated by potential function and its whole set of odd number derivatives. Such new method in this work not only contain extraction for quantum corrections in Shifted Maxwellian distribution function, but also significantly contribute the potential studying in quantum correction theory. In addition, more work is still needed to make a cumulant expansion a viable method for collision term in order to build an integrated model for quantum devices simulation.

**References**

[1] M. Müller, J. Schmalian and L. Fritz, "Graphene: a nearly perfect fluid", Physical Review Letters, **103**, 025301(1-4) (2009).

[2] J. Lee, M. Shin, S. Byun and W. Kim, "Wigner Transport Simulation of Resonant Tunneling Diodes with Auxiliary Quantum Wells", Journal of the Korean Physical Society. 72, 622-627 (2018).

[3] S. Srivastava, H. Kino, S. Nakaharai, E. Verveniotis, Y. Okawa, S. Ogawa, C. Joachim and M. Aono, "Quantum transport localization through graphene", Nanotechnology. **28**, 035703(1-10) (2017).

[4] S. Lagasse, C. Cress, H. Hughes and J. Lee, "Theory of Quantum Transport in Graphene Devices with Radiation Induced Coulomb Scatterers", IEEE Transactions on Nuclear Science. **64**, 156-163 (2017).

[5] L. Barletti, C. Negulescu, "Hybrid Classical-Quantum Models for Charge Transport in Graphene with Sharp Potentials", Physics. **9**, 7-11 (2017).

[6] A. Majorana, G. Mascali, V. Romano, "Charge transport and mobility in monolayer graphene", Journal of Mathematics in Industry. **7**, 1-13 (2016).

[7] C. Kammerer, F. Méhats, "A kinetic model for the transport of electrons in a graphene layer", Journal of Computational Physics. 327, 450-483 (2016).

[8] O. Morandi, F. Schürrer, "Wigner model for quantum transport in graphene", Journal of Physics A Mathematical & Theoretical. **44**, 2065-2088 (2011).

[9] C. Garner, "The Quantum Hydrodynamic Model for Semiconductor Devices", SIAM Journal on Applied Mathematics. **54**, 409-427 (1994).

[10] N. Zamponi and L. Barletti, "Quantum electronic transport in graphene: A kinetic and fluid-dynamic

approach", Mathematical Methods in the Applied Sciences. **34**, 807-818 (2011).

[11]A. Bose and M. Janaki, "A different approach to obtain Mayer's extension to stationary single particle Wigner distribution", Physics of Plasmas. **19**, 072101(1-4) (2012).

[12]A. Bose and M. Janaki, "A simple method to obtain the equilibrium solution of Wigner-Boltzmann equation with all higher order quantum corrections", The European Physical Journal B. 87, 259(1-6) (2014).

[13] E. Wang, M. Stettler, S. Yu and C. Maziar, "Application of Cumulant Expansion to the Modeling of Non-local Effects in Semiconductor Devices", Extended Abstracts of Sixth International Workshop on Computational Electronics, (1998).

[14] R. Kubo, "Generalized Cumulant Expansion Method", Journal of the Physical Society of Japan. 17, 1100-1120, (1962).

[15] P. Degond, "Quantum Moment Hydrodynamics and the Entropy Principle", Journal of Statistical Physics, **112**, 587-628 (2003).